\documentclass[12pt]{article}
\usepackage{epsfig}
\usepackage{graphicx}

\def\beq{\begin{equation}}
\def\eeq#1{\label{#1}\end{equation}}
\def\eeqn{\end{equation}}

\def\beqa{\begin{eqnarray}}
\def\eeqa#1{\label{#1}\end{eqnarray}}
\def\eeqan{\end{eqnarray}}

\let\bar=\overbar

\def\Dslash{\not{\hbox{\kern-4pt $D$}}}
\def\dslash{\not{\hbox{\kern-2pt $\del$}}}

\def\msb{{\bar{\ssstyle M \kern -1pt S}}}

\usepackage{fancyhdr,graphicx}
\def\Title#1{\begin{center} {\Large {\bf #1} } \end{center}}

\begin{document}

\Title{Exotic Phases in Magnetars}

\bigskip\bigskip


\begin{raggedright}

{\it 
S. Schramm$^{1}$~~A. Bhattacharyya$^{2}$~~V. Dexheimer$^{3}$~~R. Mallick$^{1}$\\
\bigskip
$^{1}$Frankfurt Institute for Advanced Studies,
Ruth-Moufang-Str. 1,
60438 Frankfurt am Main, 
Germany\\
\bigskip
$^{2}$Department of Physics,
University of Calcutta,
700009 Kolkata,
India\\
\bigskip
$^{3}$Department of Physics,
Kent State University,
Kent, OH 44242,
USA\\
}

\end{raggedright}

\section{Introduction}

The study of dense and hot matter is at the forefront of modern nuclear/heavy-ion physics and nuclear astrophysics.
While heavy-ion collisions largely probe hot matter with low net baryon density, the interior of neutron stars consists of the complementary 
phase of strongly interacting matter at extremely high densities and relatively low temperatures.
Therefore, investigating neutron star properties, apart from the interest in the objects themselves, can provide essential clues to the general properties of matter under extreme conditions. 
\\ \\
In very dense star cores, the possibility of the existence of exotic phases arises. Among them phases of Bose condensates of mesons might occur. In particular, Kaon (K$^-$) condensates have been extensively studied in the past \cite{kaplan}. Depending on the model used to study Kaon condensation in the neutron star this phase might or might not occur. One problem associated with such a system is that the condensate softens the equation of state significantly. Given the background of the measurements of 2-solar-mass neutron stars \cite{demorest,antonidis} this softening exacerbates the problem of modelling heavy stars. 
\\ \\
In addition to high densities, neutron stars also feature very strong magnetic fields of typical values around $10^{8}$ to $10^{12}$ G at the surface. In addition, a number of stars, termed magnetars have been identified to support surface magnetic fields of the order of $10^{15}$ G \cite{paczynski,duncan,thompson,melatos}.
On the other hand, very little is known about the strength of the magnetic field in the interior of the star. General estimates based on the virial theorem as well as various 2-D calculations of magnetised neutron stars point to maximum values in the core region of several times $10^{18}$ G to perhaps $10^{19}$ G \cite{bocquet,cardall,mallick}. Without addressing the question of the possible origin of such high fields, we consider field strengths up to $7\times 10^{18}$ G in the center of the neutron star to study maximum magnetic field effects.
\\ \\
The appearance of vector meson condensates due to the effect of strong magnetic fields has been proposed previously in Ref. \cite{schramm1}. The basic mechanism originates from the fact higher-spin bosons experience a strong interaction with the magnetic field leading to the potential onset of condensation. One place for such strong fields is in the central zone of a heavy-ion collision, where $\rho$ meson condensation has been considered \cite{chernodub1}. The problem in triggering condensation in such an environment comes from the necessary reduction of the meson mass to zero, leading to very large required field strengths of about $10^{20}$ G.
On the other hand, in a neutron star, negative mesons condense when their effective masses drop below the Fermi energy of the electrons, effectively replacing electrons as negatively charged particles to ensure charge neutrality. Such an effect facilitates considerably meson condensation. In this context, $\rho$ meson condensation has been considered in Refs. \cite{voskresensky,kolomeitsev}. Thus, it is an interesting task to look at the combination of both effects by studying the $\rho^-$ meson in stellar matter with large magnetic fields.

\section{Theoretical approach and results}

In order to keep the discussion simple we adopt a standard RMF type model (GM3 \cite{glendenning})  for the description of the neutron star matter.
Here, we avoid the complication of additionally considering hyperons. As done in Ref. \cite{dexheimer}, we include magnetic field effects by summing over the nucleonic and leptonic Landau levels to obtain the scalar and vector baryon densities for the equations of motion and thermodynamic quantities of the RMF solutions (for details see Ref. \cite{dexheimer}). The energy levels of the $\rho^-$ meson are given by
\begin{equation}
E_\rho = \sqrt{p_z^2 + m_\rho^2 + (2n-g_B S_z+1)eB}
\end{equation}
with $g_B = 2$, $n$ being the Landau level, and $S_z =\pm 1$ the spin projection onto the magnetic field axis (chosen to be the z-axis). 
Thus, the lowest Landau level at zero longitudinal momentum has an energy of $E_\rho^2 = m_\rho^2 -eB$. Without attempting a full survey of scaling masses (as it has been done in \cite{kolomeitsev} in the case without magnetic fields), we assume a simple field-dependent term coupling the rho meson and the scalar Walecka $\sigma$ field: 
\begin{equation}
m_\rho^* = m_\rho - g \sigma
\end{equation}
 with a free parameter $g$, thus yielding the lowest-lying $\rho^-$ state \cite{mallick2}
\begin{equation}
(E^*)_\rho = \sqrt{(m^*)_\rho^2 - eB}~~~.
\end{equation}
As the field-dependent and, therefore, density-dependent $\rho$ meson mass changes the isospin properties of the model, we readjust the nucleon-$\rho$ coupling $g_{N\rho}$ for each different choice of $g$ in order to keep the asymmetry parameter at its original value in Ref. \cite{glendenning} of $a_{sym} = 32.5 $ MeV.

In order to estimate whether, in principle, $\rho^-$ condensation is can occur in a neutron star, we choose a high magnetic field value $B = 7 \times 10^{18}$ G. Note that this value is at the upper end of the hypothetically possible values for the field at the center of neutron stars. Fig. 1 shows the results for $(E^*)_\rho$ for different values of $g$ as function of baryon density. As expected, $E^*$ drops with density, while the Fermi energy of the electron increases. As can be read off the plot, both lines cross at a density around $4.4\times \rho_0$ for a coupling $g = - 6$. Using these parameters, such a central density is reached only for stellar masses above $2\,M_\odot$. As in this case the maximum mass without condensation reaches $M_{max} = 2.13\,M_\odot $, it follows that there is a range of possible stars with condensation. Note that n this simplified investigation we have not yet considered the effects of the condensate on the stellar solution after condensation sets in, but rather only determined the critical densities for its appearance. The most drastic effect could be a cut-off of stable stars at the onset of condensation, due to the softening of the equation of state.
\begin{figure}[htb]
\includegraphics[width=0.6\textwidth]{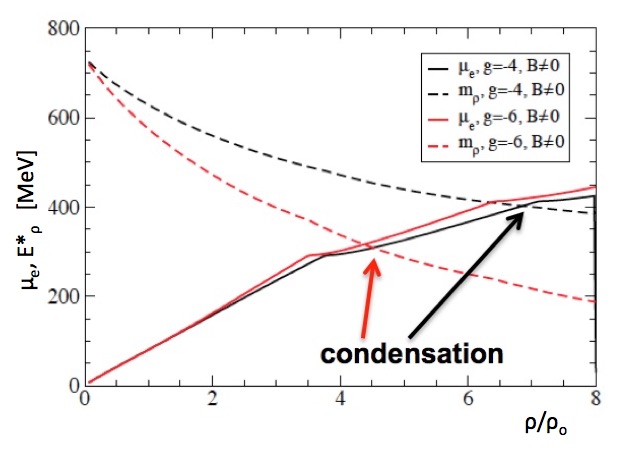}
\caption{Effective energy of $\rho^-$ meson and electron fermi energy as functions of normalized baryon density. Different choices for the parameter $g$ are compared. The crossing of the curves mark the onset of condensation.}
\label{cond}
\end{figure}

One additional point has to be considered when introducing a density dependence of the $\rho$ meson mass. While we readjusted  $g_{N\rho}$ in order to keep $a_{sym}$ fixed, the value of the slope parameter $L = 3\, \rho_0\, [da_{sym}/d\rho](\rho_0)$ changes with the parameter $g$. This is illustrated in Fig. 2.
Here, the value of $L$ is shown  as function of the critical density of $\rho$ condensation, which corresponds to different choices of the coupling $g$. As one can see, an onset of condensation at about $5\times \rho_0$ corresponds to an increase of  20 MeV in the value of $L$ compared to the original model. This,  in turn, restricts the possible range of values of $g$.
\\ \\
While our investigations have been performed using the GM3 parameterization, the actual quantitative results might shift in different models. Still, the overall qualitative results should remain unchanged. Additionally, it would be interesting to investigate possible consequences of a $\rho$ meson condensate on processes like neutron star cooling. Note that, even if the condensate leads to a cut-off of possible neutron star masses below the phenomenological 2 solar mass value, this would not be in conflict with observation, as such a mass-reduction only takes place in highly magnetized stars.


\begin{figure}[htb]
\includegraphics[width=0.6\textwidth]{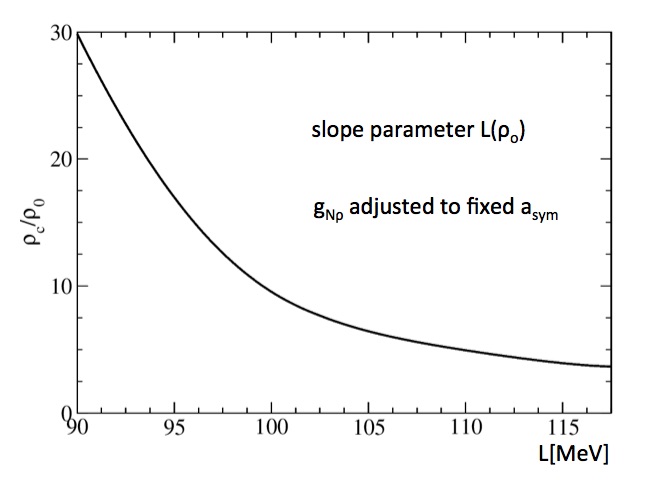}
\caption{Slope parameter L as function of the critical density for $\rho^-$ condensation.}
\label{slope}
\end{figure}


\subsection*{Acknowledgement}

We thank the organizers of CSQCDIV for a very enjoyable and fruitful meeting.
SWS acknowledges support from the LOEWE program HIC for FAIR.

\end{document}